\documentclass{icrc2009}

\usepackage{graphicx}   
\usepackage{caption}    
\usepackage[font=footnotesize]{subfig} 
\usepackage{fixltx2e}
\usepackage{url}

\newcommand{\shorttitle}[1]%
{\markboth{Proceedings of the 31\MakeLowercase{$^{st}$} ICRC,
{\L}\'{o}d\'{z} 2009}{#1} }
\newcommand{\etal}{\MakeLowercase{\textit{et al. }}} 

\begin{document}
\title{MAGIC observation of Globular Cluster M13 and its millisecond pulsars}

\author{\IEEEauthorblockN{T.~Jogler\IEEEauthorrefmark{1},
              C.~Delgado Mendez\IEEEauthorrefmark{2}\IEEEauthorrefmark{3},
                          M.~T.~Costado\IEEEauthorrefmark{2}\IEEEauthorrefmark{4},
                           W.~Bednarek\IEEEauthorrefmark{5} and
                           J.~Sitarek\IEEEauthorrefmark{1}\IEEEauthorrefmark{5}}
                           on behalf of the MAGIC collaboration
                            \\
\IEEEauthorblockA{\IEEEauthorrefmark{1}Max-Planck-Institut f\"ur
Physik, D-80805 M\"unchen,Germany}
\IEEEauthorblockA{\IEEEauthorrefmark{2}Inst. de Astrof\'{\i}sica
de Canarias, E-38200 La Laguna, Tenerife, Spain}
\IEEEauthorblockA{\IEEEauthorrefmark{3}now at: CIEMAT, Avda.
Complutense 22, E-28040, Madrid, Spain}
\IEEEauthorblockA{\IEEEauthorrefmark{4}Depto. de Astrofisica,
Universidad, E-38206 La Laguna, Tenerife, Spain}

\IEEEauthorblockA{\IEEEauthorrefmark{5}University of \L\'od\'z,
PL-90236 Lodz, Poland}}

\shorttitle{T.~Jogler \etal MAGIC observation of M13} \maketitle

\begin{abstract}
Based on MAGIC observations from June and July 2007, we present
upper limits to the $E>140~\textrm{GeV}$ emission from the
globular cluster M13. Those limits allow us to constrain the
population of millisecond pulsars within M13 and to test models
for acceleration of leptons inside their magnetospheres and/or
surrounding. We conclude that in M13 either millisecond pulsars
are fewer than expected or they accelerate leptons less
efficiently than predicted.
  \end{abstract}


\section{Introduction}
Globular clusters (GC) are very interesting sites for probing high
energy processes due to their large content of evolved objects
like millisecond pulsars (MSP). It has been estimated that a
typical massive GC contains of the order of 100 MSP
\cite{tavani2}. Up to now the largest samples of MSPs discovered
are located in the GC Ter 5 (23 MSP), and Tuc 47 (22 MSP) (see
e.g. \cite{camilo}).

TeV $\gamma$-rays fluxes from GC have been predicted based on
estimates on the population of MSPs and the efficiency of lepton
acceleration in their surrounding (see \cite{wlolek} and
\cite{venter}). The flux level is detectable by the current
generation of cherenkov telescopes like MAGIC. The $\gamma$-rays
would be produced by accelerated leptons scattering off photons of
the microwave background radiation or the thermal emission of an
extremely dense cluster of solar mass stars inside the GC.
Acceleration of leptons could take place either in the shocks
 produced by the collision of the MSPs winds within the GC, or in
the pulsar inner magnetosphere or their wind regions. Furthermore,
the inner MSP magnetosphere could be a production site for
$\gamma$-rays in the sub-TeV energy range as it is predicted from
model calculations~\cite{bulik,harding}. Additional contribution
to the sub-TeV $\gamma$-ray emission could be produced in the
vicinity of radio emitting blocked pulsars \cite{aharonian,albert
s} inside low mass binary systems \cite{tavani91}.

No VHE $\gamma$-ray emission could be detected so far in any GC.
The few experimental results on VHE $\gamma$-ray emission reported
in the literature are upper limits on the emission of M13 by the
{\it WHIPPLE} Collaboration \cite{hall}, M15 by the {\it VERITAS}
Collaboration \cite{lebohec}, and $\omega$ Centauri by the {\it
CANGAROO} Collaboration \cite{kabuki}. Very recently, the {\it
Fermi} LAT telescope has detected high energy $\gamma$-ray
emission ($E>100~\textrm{MeV}$) from one of the closest and most
massive GC, Tuc 47 \cite{guillemot}, and {\it HESS} has obtained
an upper limit of
$6.7\times10^{-13}~\textrm{ph}~\textrm{cm}^{-2}~\textrm{s}^{-1}$
for energies $E>800~\textrm{GeV}$ \cite{hess tuc}, but given the
possible complexity of the emission in the GeV range, it is not
possible to establish any connection between these results. From
the {\it HESS} result constrains to the magnetic field in the
pulsar nebula as a function of the number of MSP in the GC can be
drawn for the model in~\cite{venter}. In addition restrictions to
the efficiency of the rotational energy converted by the MSPs into
relativistic leptons can be placed for the model described in
\cite{wlolek}.

Here we report the results of observations with the MAGIC
telescope of the globular cluster M13, and present the constraints
that our results impose to the population of millisecond pulsars
and their lepton acceleration efficiency. M13 belongs to the class
of ordinary globular clusters, and its estimated mass is
$6\times10^5~\textrm{M}_\odot$. It is located in the northern
constellation Hercules at a distance of $~7$ kpc and thus one of
the closer GC. Its core radius is about $\sim1.6$ pc, with a half
mass radius of $\sim3.05$ pc \cite{harris}. Up to now 5
millisecond pulsars have been detected in M13, with periods
ranging between 2 and 10 ms. The {\it WHIPPLE} Collaboration
derived from their observation a flux upper limit of $1.08\times
10^{-11}~\textrm{ph.}~\textrm{cm}^{-2}~\textrm{s}^{-1}$ at
energies $E>500~\textrm{GeV}$~\cite{hall}.

\section{Observations and data analysis}
The MAGIC telescope is an Imaging Atmospheric Cherenkov Telescope
(IACT) located at the Observatory Roque de los Muchachos on the
Canarian Island La Palma ($28.75^\circ$N, $17.86^\circ$W, 2225~m
a.s.l.). It has an exceptional light detection efficiency provided
by the combination of a 17 m diameter mirror and a pixelized
camera composed of 576 high quantum efficiency, hemispherical
photomultiplier tubes (PMT). The standard trigger threshold of
MAGIC is $\sim{60}~\textrm{GeV}$. For energies above
$150~\textrm{GeV}$, the telescope angular and energy resolutions
are $\sim{0.1}~\textrm{deg.}$ and $\sim{25}$\% respectively (see
\cite{albert} for further details). In April 2007 the data
acquisition system of MAGIC was upgraded with multiplexed 2 GHz
Flash Analog-Digital converters which improved the timing
resolution of the recorded shower images~\cite{mux}. Accordingly
the sensitivity of MAGIC improved significantly  to 1.6\% of the
Crab Nebula flux above $270~\textrm{GeV}$ for 50 hours of
observation~\cite{aliu}.

We observed M13 at zenith angles ranging from $8^{\circ}$ to
$31^{\circ}$ between June 12th and July 18th of 2007 in the
false-source tracking (wobble) mode \cite{fomin}. Two tracking
positions 24' offset in RA on opposite sides of M13 were used.
This technique allows for a reliable estimation of the background
with no need of extra observation time. The collected data amount
to 20.7 hours effective observation time after rejecting events
affected by unstable hardware or environmental conditions. Besides
this, events with a collected charge below 300 photo-electrons
were rejected in order to maximize the analysis sensitivity. Due
to this selection the analysis threshold is 190~GeV.

The data analysis was carried out using the standard MAGIC
analysis and reconstruction software chain, which proceeds in
several steps. After the standard calibration of the PMT signal
pulses~\cite{albert a}, pixels containing no useful information
for the shower image reconstruction are discarded by an image
cleaning procedure~\cite{aliu}. Then for the events image
parameters are calculated \cite{hillas} using the surviving
pixels. In addition to the classical Hillas parameters, two timing
parameters are computed, namely: the gradient of the arrival times
of the Cherenkov photons along the shower axis; and their arrival
time spread over the whole shower. The primary particle
identification in each event is achieved using a multidimensional
classification procedure based on the Random Forest (RF) method
\cite{albert b}. In the RF the probability to be a hadron induced
event, the so called hadronness, is computed for each event based
on its image and time parameters. Another RF is trained with a
Monte Carlo simulated $\gamma$-ray sample to estimate the energy
on an event by event basis. Finally the angle between the major
axis of the shower image ellipse and the source position in the
camera, the so called Alpha angle, is used to select $\gamma$-ray
candidates in the direction of the source. To estimate the
remaining background, the angle Alpha is also computed with
respect to the anti source position. The anti source position is
$180^{\circ}$ rotated to the source position with respect to the
camera center.

Main contributions to the systematic uncertainties of our analysis
are the uncertainties in the atmospheric transmission, the
reflectivity (including stray-light losses) of the mirrors and the
light catchers, the photon to photo-electron conversion
calibration and the photo-electron collection efficiency in the
photomultiplier front-end. A detailed discussion of their
contribution to the flux uncertainties can be found in
\cite{albert}, where they are estimated to add up to 30\% of the
measured flux value.

\section{Results}

\begin{figure}[b]
\includegraphics[width=\linewidth]{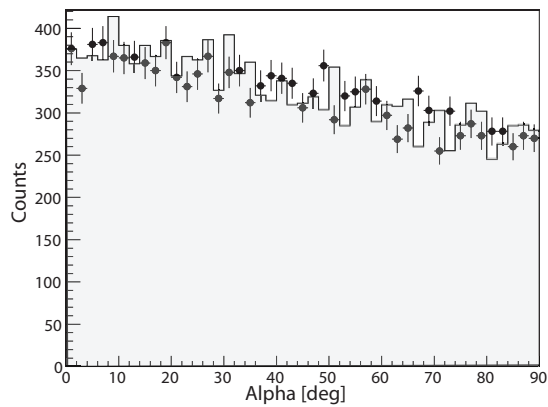}
\caption{The Alpha distribution for the selected $\gamma$-ray
candidates from the signal (black dots) and background (grey
shaded) region. No significant signal is present in the data
sample. }\label{fig:alpha}
\end{figure}

Figure \ref{fig:alpha} shows the obtained Alpha angle distribution
for the source and the background regions. A hadronness cut tuned
to yield an energy independent $\gamma$-ray selection efficiency
of 80\%, estimated by means of a Monte Carlo simulation has been
applied. We define the signal region as the smaller interval in
Alpha angle that contains 80\% of the $\gamma$-rays for each
energy bin, estimated using a Monte Carlo simulation. Their lower
bounds are at Alpha$=0$ and the upper ones are shown in second
column of Table \ref{tab:uldif} for each energy bin. We find
$-23\pm57$ excess events after background subtraction in the
signal region for energies above $E=140\textrm{ GeV}$.
Furthermore, no significant signal is present in a region
extending $1~\textrm{deg}$ of radius around M13. The obtained
upper limits to the VHE flux from M13 for different energy bins,
are shown in Table \ref{tab:uldif}. These have been computed using
the Rolke method~\cite{rolke} at a $95\%$ confidence level, and
they take into account a $30\%$ systematic uncertainty in the flux
level. The upper limit to the integral flux for energies above
$E=200~\textrm{GeV}$, assuming a spectral index of 2.6, is
$5.1\times10^{-12}~\textrm{cm}^{-2}~\textrm{s}^{-1}$.

\begin{table}[h]
\caption{Differential upper limits} \vspace*{-0.3cm}
\begin{center}
\resizebox{\linewidth}{!}{
\begin{tabular}{cccccc}
\hline \hline
Energy bin  & Upper Alpha & Events & Background& Excess UL & Flux UL \\
GeV         & cut (deg)   &        & events    &(95\% CL)& (cm$^{-2}$ s$^{-1}$TeV$^{-1}$)  \\
\hline
$140~-~200$ &8 & $487$  & $517\pm23$   &37   &7.2 $\times$ 10$^{-11}$ \\
$200~-~280$ &10 & $683$  & $681\pm27$   &95   &5.1 $\times$ 10$^{-11}$\\
$280~-~400$ &8 & $254$  & $242\pm16$   &75   &2.2 $\times$ 10$^{-11}$\\
$400~-~560$ &6 & $62$   & $73\pm9$     &14   &2.4 $\times$ 10$^{-12}$\\
$560~-~790$ &4 & $32$   & $27\pm5$     &27   &2.7 $\times$ 10$^{-12}$\\
$790~-~1120$ &4 & $4$    & $5.7\pm2.4$  &5.8  &3.7 $\times$ 10$^{-13}$\\
 \hline
\end{tabular}
}
\end{center}
\label{tab:uldif}
\end{table}

\begin{figure}
\includegraphics[width=\linewidth]{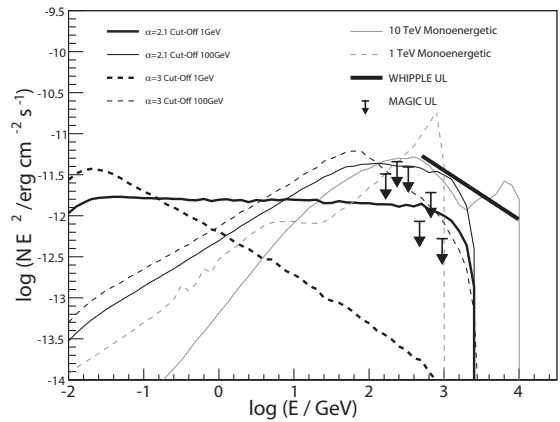}
\caption{ The MAGIC $\gamma$-ray flux upper limits for M13
compared with spectra expected for the range of parameters of the
model shown in Figure 9 and 10 of~\cite{wlolek}. The specific
$\gamma$-ray spectra are calculated for lepton upper energy
cut-off at 3 TeV and lower energy cut-off at 1 GeV (black thick)
and 100 GeV (black thin), and power-law spectral indices of 2.1
(solid) and 3 (dashed). The $\gamma$-ray spectra produced by
mono-energetic leptons of 10 TeV and 1 TeV are shown by a grey
solid curve and a dashed one respectively. All calculations are
computed assuming the conservative value of $1$ for the free
parameter of the model $N_{\rm MSP}\cdot \eta$. The Whipple
differential upper limit shown here has been derived from the
integral quoted in~\cite{hall} assuming a spectral index of 2.6.}
\label{fig:ul}
\end{figure}

\section{Comparison with models}
In Figure~\ref{fig:ul} we compare our flux upper limits with the
theoretical $\gamma$-ray spectra calculated in~\cite{wlolek}. In
this model, leptons are injected into the GC volume according to a
power-law spectrum, upon acceleration in the shocks produced in
the collisions of the pulsar winds of several MSPs. The
$\gamma$-rays are then produced via Inverse Compton scattering of
 microwave background radiation photons and from thermal
radiation arising from the whole GC. Constrains to the total power
of all injected leptons ($L_{\rm e}$) can be derived by comparing
our upper limits to the theoretical expected values. The
theoretical predictions are required to be always lower than our
upper limits and the we assume a distance of 7~kpc to M13. The
thus derived upper limits to $L_{\rm e}$ are reported in
Table~\ref{tab1} for different assumptions of the spectral shape
of the injected leptons, i.e. for different values of the spectral
index $\alpha$ between the minimum energy $E_{\rm min}$ and the
maximum energy defined by the escape of leptons from the shock. We
calculate the upper limits for two energies in the case of mono
energetic lepton injection (1~TeV and 10~TeV). Generic parameter
values for MSPs (surface magnetic field $10^9$ G and rotational
period 4 ms) are assumed. This enables us to translate the limits
on $L_{\rm e}$ into limits to the product of the number of MSPs in
M13 ($N_{\rm MSP}$) times the efficiency of the rotational energy
conversion of MSPs into relativistic leptons ($\eta$). The results
are shown in Table~\ref{tab1}. In example up to 100 MSPs are
predicted to be contained in M13~\cite{tavani2}. On the other
hand,
 the efficiency of lepton injection from the inner magnetospheres of
millisecond pulsars has been estimated to be $\eta\sim 0.1$ in
terms of the extended polar gap model~\cite{muslinov}. Hence the
product of $N_{\rm MSP}\cdot \eta$, can be most likely of the
order of $\sim 10$. In Table~\ref{tab1} our upper limits to this
product for the various considered model parameters are shown. For
most of the considered models $N_{\rm MSP}\cdot \eta$ is
significantly below $\sim 10$. The model with the soft spectrum of
leptons which extends down to 1 GeV is the only one we can not
strongly constrain. Note that even if the number of MSP in M13 is
only equal to 5 (which is the up to now detected
number~\cite{camilo}), we can already obtain the acceleration
efficiency of leptons to be $\sim$0.1 in the case of their
injection with the hard (spectral index 2.1) and mono-energetic
spectrum respectively.

\begin{table}
\caption{Upper limits on the power of injected leptons ($L_{\rm
e}$) and in $N_{\rm MSP}\cdot \eta$} \label{tab1}
\resizebox{\linewidth}{!}{
\begin{tabular}{c c c c c c c}
\hline\hline
$E_{\rm min}$ &100 GeV&100 GeV&1 GeV&1 GeV&mono:&mono:   \\
 $\alpha$     & 2.1   & 3.0   & 2.1 & 3.0 &1 TeV&10 TeV  \\
\hline
              &       &       &     &     &     &  \\
$L_{\rm e}$   &0.6 & 1.0& 1.0& 60& 0.2& 0.5  \\
\footnotesize{($\times10^{35}~\textrm{erg}~\textrm{s}^{-1}$)}  &    &    &    &    &    &  \\
                      &    &    &    &    &    &  \\
$N_{\rm MSP}\cdot \eta$ &0.5&1.0 &  1.0 &50 &0.2&  0.4 \\
                      &    &    &    &    &    &  \\
\hline\hline
\end{tabular}
}
\end{table}

\section{Conclusions}

We present the strongest upper limits to date on the VHE
$\gamma$-ray flux from the massive globular cluster M13. Our upper
limit is $\sim 2$ times lower than the previous reported limit for
VHE energy emission from M13 quoted by {\it WHIPPLE}, and extends
to energies down to 140 GeV. With these upper limits we can
constrain the population of the millisecond pulsars expected in
M13 and the acceleration scenarios of leptons by millisecond
pulsars. Our result strongly suggests that either the number of
millisecond pulsars in M13 is significantly lower than the
estimate of $\sim 100$, or the energy conversion efficiency from
millisecond pulsars to relativistic leptons is significantly below
the value quoted in recent modeling of high energy processes in
the magnetospheres of millisecond pulsars.

\section*{Acknowledgements}
We would like to thank the Instituto de Astrofisica de Canarias
for the excellent working conditions at the Observatorio del Roque
de los Muchachos in La Palma. The support of the German BMBF and
MPG, the Italian INFN and Spanish MICINN is gratefully
acknowledged. This work was also supported by ETH Research Grant
TH 34/043, by the Polish MNiSzW Grant N N203 390834, and by the
YIP of the Helmholtz Gemeinschaft.

\end{document}